\documentclass[11pt, letterpaper]{elsarticle}

\usepackage{amsmath}
\usepackage{amssymb}
\usepackage{amsfonts}
\usepackage{graphicx}
\usepackage{multirow}
\usepackage{latexsym}
\usepackage{epic}
\usepackage{url}
\usepackage{soul}
\usepackage{afterpage}
\usepackage[unicode, pdftex]{hyperref}
\usepackage{multicol}
\usepackage{bbm}
\usepackage{setspace}
\usepackage{enumitem}
\usepackage{cleveref}
\usepackage[toc]{appendix}
\usepackage[norelsize,ruled, lined,linesnumbered]{algorithm2e}
\usepackage{amsthm}
\usepackage[symbol]{footmisc}
\usepackage[mathscr]{euscript}
\usepackage[left=0.9in,top=0.9in,right=0.9in,bottom=0.9in,nohead]{geometry}
\usepackage{empheq}
\usepackage{mathtools}
\DeclareMathOperator{\argmax}{argmax}
\DeclareMathOperator{\argmin}{argmin}

\makeatletter
\newtheorem*{rep@theorem}{\rep@title}
\newcommand{\newreptheorem}[2]{%
	\newenvironment{rep#1}[1]{%
		\def\rep@title{#2 \ref{##1}}%
		\begin{rep@theorem}}%
		{\end{rep@theorem}}}
\makeatother
\SetAlgoNlRelativeSize{-1}
\SetKwFor{For}{for}{}{end}
\SetKwFor{While}{while}{}{end}
\newreptheorem{example}{Example}
\newtheorem{theorem}{Theorem}

\newtheorem{lemma}{Lemma}
\newtheorem{proposition}{Proposition}

\newtheorem{corollary}{Corollary}

\usepackage{xcolor}
\definecolor{forestgreen}{RGB}{34,139,34}
\usepackage[colorinlistoftodos,prependcaption,textsize=tiny]{todonotes}
\usepackage{xargs}
\usepackage{booktabs}
\usepackage{caption}
\usepackage{subcaption}

\usepackage{lineno}
\usepackage{float}
\usepackage[section]{placeins}
\usepackage{arydshln}
\usepackage{latexsym}
\usepackage{epic}
\usepackage{amsmath,amsthm,amssymb}
\usepackage{graphicx}
\usepackage{url}
\usepackage{pgfplots}\usetikzlibrary{plotmarks}
\usepackage{soul}
\usepackage{afterpage}
\usepackage{cleveref}
\usepackage{bbm}
\usetikzlibrary{decorations.markings}
\usepackage{setspace}
\usepackage{appendix}
\usepackage{comment}
\DeclareGraphicsExtensions{.pdf,.png,.jpg,.bmp}

\allowdisplaybreaks
\makeatletter
\def\ps@pprintTitle{%
	\let\@oddhead\@empty
	\let\@evenhead\@empty
	\let\@oddfoot\@empty
	\let\@evenfoot\@oddfoot}
\makeatother

\usepackage{etoolbox}

\begin{document}
	\begin{frontmatter}
		\title{	\doublespacing On the Complexity of Bilevel Linear and Quadratic Programs \\ in Fixed Dimensions}
		\author[label1]{Sergey S.~Ketkov\footnote[2]{Corresponding author. Email: sergei.ketkov@business.uzh.ch; phone: +41 078 301 85 21.}}
		\author[label1]{Oleg A. Prokopyev}
		\address[label1]{Department of Business Administration, University of Zurich, Zurich, 8032, Switzerland}
		\begin{abstract} 	\doublespacing
	 We study the computational complexity of bilevel linear programs (BLPs) and bilevel quadratic programs (BQPs) with continuous variables under restrictions on the numbers of variables or constraints. 
	 Our results identify new transitions between computational complexity regimes that are inherent to the hierarchical structure of bilevel optimization problems and have no direct counterparts in single-level optimization.  
    First, we demonstrate that, under comparable assumptions, the \textit{pessimistic} formulation can be computationally harder than its \textit{optimistic} counterpart.  Specifically,  while optimistic BLPs are known to be polynomially solvable when the number of follower \textit{variables} is fixed, we prove that the corresponding pessimistic problem is strongly~$\mathrm{NP}$-complete.   We also show that this complexity transition is tied to the presence of a nonfixed number of \textit{coupling constraints}.  Second, we show that, in the pessimistic setting with a fixed number of follower \textit{constraints}, BLPs are polynomially solvable, whereas replacing the follower's linear objective function with a convex quadratic one renders the problem $\mathrm{NP}$-hard. Importantly, this complexity transition 
    occurs even in the absence of coupling constraints.  Finally, we demonstrate that replacing a convex quadratic follower objective with a nonconvex quadratic one renders the optimistic problem $\mathrm{NP}$-hard, even when both follower dimensions are fixed  and~small.

		\end{abstract}
		
		
	\end{frontmatter}

\section{Introduction} \label{sec: intro} \doublespacing
Identifying transitions between computational complexity regimes is fundamental to understanding how the structure of an optimization problem determines its computational difficulty.  Such transitions can arise from restrictions on the numbers of \textit{variables} or \textit{constraints}, as well as from changes in the \textit{form} of the objective function.  A classical example is integer linear programming, which is $\mathrm{NP}$-hard in general but polynomial-time solvable when the number of integer \textit{variables} is fixed~\cite{Lenstra1983}.  Another example involves polynomial optimization:  quadratic optimization with a fixed number of quadratic constraints is polynomial-time solvable~\cite{Grigoriev2005}, whereas minimizing a cubic polynomial with a single convex quadratic constraint is $\mathrm{NP}$-hard~\cite{Nesterov2003}.

This note studies analogous complexity transitions in \textit{bilevel optimization}. In contrast to single-level optimization, a bilevel problem  models a hierarchical interaction between two decision-makers, commonly referred to as a \textit{leader} and a \textit{follower}, each with  their  own objective function and constraints. The leader makes a decision first, optimizing  their  objective function and anticipating that the follower subsequently solves  their  own optimization problem, parameterized by the leader's decision; see, e.g., \cite{Colson2007, Kleinert2021,Sinha2017} for comprehensive surveys of bilevel problems and their applications.

Several distinctive features of the bilevel setting are important for our analysis. First, fixing the number of \textit{variables} or \textit{constraints} for the leader or the follower can have different complexity implications. Second, when the follower has multiple optimal solutions, the \textit{optimistic} and \textit{pessimistic} formulations assume favorable and unfavorable selection, respectively, from the leader's perspective. Finally, bilevel models may also include \textit{coupling constraints}. Their presence allows the follower's optimal decision to affect not only the leader's objective function but also the leader's feasible set. 

These modeling features also have direct practical significance. In particular, dimensional restrictions naturally arise in applications where one decision-maker controls only a few variables or constraints. For example, in bilevel continuous knapsack problems, the leader may only determine the  follower's  total knapsack capacity, whereas in robust portfolio optimization the follower may only control a return threshold; see, e.g., \cite{Buchheim2022, Conn2012}.
Furthermore, the distinction between optimistic and pessimistic formulations arises in toll pricing, where a network operator (leader) sets tolls while 
travelers (followers) choose minimum-cost routes, with equally optimal routes generating different toll revenues~\cite{Labbe1998,Roch2005}.
Finally, related network-pricing models illustrate the role of coupling constraints. In such models, a regulator (leader) chooses transport subsidies while mobility service providers (followers) set fares, with an upper-level budget constraint linking the regulator's subsidy decisions to the demand induced by the followers' optimal fares; see, e.g.,~\cite{Huang2024}.


\looseness-1 It is known that general \textit{bilevel linear programs} (BLPs), without restrictions on the numbers of variables or constraints, are strongly $\mathrm{NP}$-hard~\cite{Hansen1992}. Motivated by the structural features discussed above, we therefore study how the computational complexity of BLPs 
is affected by fixing the number of variables or constraints at either level, adopting optimistic or pessimistic conventions, and allowing for coupling constraints.  We first complete the complexity classification of BLPs in fixed dimensions and then investigate how this complexity landscape changes  for \textit{bilevel quadratic programs} (BQPs) obtained by generalizing the decision-makers' objective functions from linear to~quadratic.

\subsection{Bilevel Linear Programs} 
\textit{Bilevel linear programs} (BLPs) constitute a canonical class of bilevel optimization problems, in which both the leader and the follower solve linear optimization problems; see, e.g., \cite{Audet1997, Hansen1992} and the survey in~\cite{Kleinert2021}. Formally, BLPs are defined as: 
\begin{subequations}
	\label{BLP}
	\begin{align}
		[\textbf{BLP}]: \quad
		\min_{\mathbf{x}} \; & 
		\mathbf{c}_{l}^{\top}\mathbf{x} + 
		\mathbf{d}_{l}^{\top}\mathbf{y}^* 
 \label{obj: leader} \\
		\text{s.t. } 
		& \mathbf{x} \in X(\mathbf{y}^*) \label{cons: leader} \\
		& \mathbf{y}^* \in 
		\argmin_{\;\mathbf{y} \in Y(\mathbf{x})} \,
		\mathbf{d}_{f}^{\top}\mathbf{y}, \label{cons: follower}
	\end{align}
\end{subequations}
Here, the leader's and the follower’s feasible sets are defined, respectively, as
\begin{subequations}
	\begin{align} 
		& X(\mathbf{y}^*) := \left\{
		\mathbf{x} \in \mathbb{R}^{n_l}_{+}
		:\;
		\mathbf{A}_l \,\mathbf{x} + 
		\mathbf{G}_l \, \mathbf{y}^* 
		\leq 
		\mathbf{h}_{l}
		\right\}, \label{eq: leader's feasible set} \\
		& Y(\mathbf{x}) = 
		\left\{
		\mathbf{y} \in \mathbb{R}^{n_f}_{+} 
		:\;
		\mathbf{A}_f \, \mathbf{x} + 
		\mathbf{G}_f \, \mathbf{y} 
		\leq
		\mathbf{h}_{f}
		\right\}, \label{eq: follower's feasible set}
	\end{align}	
\end{subequations}
 with $\mathbf{A}_i \in \mathbb{Q}^{m_i \times n_l}$,
$\mathbf{G}_i \in \mathbb{Q}^{m_i \times n_f}$,
$\mathbf{h}_i \in \mathbb{Q}^{m_i}$ for $i \in \{l,f\}$. Also, $\mathbf{c}_l \in \mathbb{Q}^{n_l}$,
$\mathbf{d}_l \in \mathbb{Q}^{n_f}$ and~$\mathbf{d}_f \in \mathbb{Q}^{n_f}$.
That is, $n_i$ and $m_i$, $i \in \{l, f\}$, correspond to the numbers of \textit{variables} and \textit{constraints}  (excluding nonnegativity constraints)  for the leader and the follower,~respectively.

With a slight abuse of notation, it is not explicitly specified in~[\textbf{BLP}] how the case of
multiple optimal solutions to the follower’s problem in \eqref{cons: follower} is addressed. Formally, the \textit{optimistic} version selects, among all follower-optimal responses~$\mathbf{y}^*$, the one that minimizes the leader’s objective function~(\ref{obj: leader}) and satisfies the \textit{coupling constraints} (\ref{cons: leader}). In contrast, the \textit{pessimistic} version guards the leader against the worst follower-optimal response $\mathbf{y}^*$ and requires constraints (\ref{cons: leader}) to hold for all follower-optimal responses; see Section \ref{subsec: blp model} for details. 
In addition, we say that [\textbf{BLP}] has no coupling constraints when~$\mathbf{G}_l = \mathbf{0}$, and it is \textit{independent} when~$\mathbf{A}_f = \mathbf{0}$; see, e.g.,~\cite{Wiesemann2013}.


It is known that [\textbf{BLP}] is strongly $\mathrm{NP}$-hard, even when $\mathbf{G}_l = \mathbf{0}$, $\mathbf{c}_l = \mathbf{0}$, and the objective functions of the leader and the follower are exact opposites, i.e., $\mathbf{d}_f = -\mathbf{d}_l$; see \cite{Hansen1992}. 
Moreover, the recent study in~\cite{Sugishita2026} demonstrates that [\textbf{BLP}] remains $\mathrm{NP}$-hard, even when the leader has a single decision variable~($n_l = 1$).
Notably, the reduction in \cite{Sugishita2026} also applies to the pessimistic formulation.  As a side observation, we show in Section \ref{subsec: bqp np} that the same result for optimistic BLPs can be obtained (with a substantially simpler proof) by leveraging complexity results for the parametric minimum-cost flow problem from~\cite{Disser2018}. 


On the other hand, some classes of BLPs are solvable in polynomial time. First, it is known that the optimistic version of [\textbf{BLP}] is polynomially solvable when the number of follower decision variables~$n_f$ is fixed \cite{Deng1998, Liu1995}. 
 Furthermore, it is shown in \cite{Wiesemann2013} that both optimistic and pessimistic \textit{independent}~BLPs admit a polynomial-time solution algorithm. 

\begin{table}[h!]
	\centering
	\setstretch{1.25}
	\renewcommand{\arraystretch}{1.2}
	\scalebox{0.95}{\begin{tabular}{@{}llll@{}}
		\toprule
		& Fixed parameter & Optimistic version & Pessimistic version \\
		\midrule
		\multirow{2}{*}{Leader} 
		& $n_l$ & \multirow{2}{*}{$\left.\vcenter{\hbox{\rule{0pt}{2.7em}}}\right\}$ $\mathrm{NP}$-complete~\cite{Disser2018, Sugishita2026}} 
		& \multirow{2}{*}{$\left.\vcenter{\hbox{\rule{0pt}{2.7em}}}\right\}$ $\mathrm{NP}$-complete~\cite{Sugishita2026}} \\[2pt]
		& $m_l$ & & \\[4pt]
		\midrule
		\multirow{2}{*}{Follower} 
		& $n_f$ & $\mathrm{P}$~\cite{Deng1998, Liu1995} & $\boldsymbol{\mathrm{NP}}$\textbf{-complete} (Theorem \ref{theorem 3}) \\[2pt]
		& $m_f$ & $\boldsymbol{\mathrm{P}}$ (Theorem \ref{theorem 1}) & $\boldsymbol{\mathrm{P}}$ (Theorem \ref{theorem 2}) \\[2pt]
		\bottomrule
	\end{tabular}}
	\caption{Computational complexity of \textbf{[BLP]} in fixed dimensions. 
	The symbol $\mathrm{P}$ denotes polynomial-time solvability, and $\mathrm{NP}$-completeness  is understood via the standard threshold decision version of [\textbf{BLP}]. Bold entries indicate our~results.}
	\label{tab:complexity}
\end{table}

In this note, we complete the complexity classification of BLPs in fixed dimensions (see Table~\ref{tab:complexity}). Our main results can be summarized as follows:
\begin{itemize}
	\item Using a value-function reformulation, we prove that the optimistic version of [\textbf{BLP}] is polynomially solvable when the number of follower constraints~$m_f$ is fixed.
	\item Using structural results from computational geometry, we show that the pessimistic version of~[\textbf{BLP}] remains polynomially solvable when the number of follower constraints~$m_f$ is~fixed.
	\item In contrast, by representing binary variables via pessimistic coupling constraints, we demonstrate that the pessimistic problem with a fixed number of follower variables $n_f$ is strongly $\mathrm{NP}$-complete.
\end{itemize}

It is worth mentioning that our conclusions differ from those reported in Deng~\cite{Deng1998}, where the analysis for a fixed number of follower variables is stated to apply to both the optimistic and pessimistic variants of~[\textbf{BLP}]. We note, however, that Deng~\cite{Deng1998} does not provide an explicit formal definition of the pessimistic formulation, and different interpretations exist in the literature, particularly regarding the treatment of coupling constraints (\ref{cons: leader}). Furthermore, Theorem~2.2 in~\cite{Wiesemann2013} builds  upon  the results of~\cite{Deng1998} and states that the pessimistic problem with coupling constraints is polynomially solvable when the number of follower variables is fixed. 
As our findings show, this is not the case~in~general; see the corresponding entries in Table \ref{tab:complexity}.

More importantly, it has long been believed that pessimistic bilevel problems are computationally more challenging than their optimistic counterparts. However, recent results by Zeng~\cite{Zeng2020} and Henke~et~al.~\cite{Henke2026} suggest that pessimistic problems with coupling constraints can, in general, be solved by leveraging their optimistic reformulations. Our results reveal the limits of this approach: the same  class of  bilevel linear programs can be polynomially solvable in the optimistic problem setting, while being strongly $\mathrm{NP}$-complete  in the pessimistic~case. Moreover, we demonstrate that the  hardness  of pessimistic BLPs is tied to the presence of  a nonfixed number of  coupling~constraints.

\subsection{Bilevel Quadratic Programs} 
Given the complexity results for BLPs in Table~\ref{tab:complexity}, a natural question is whether the polynomially solvable cases of [\textbf{BLP}] persist when the follower's objective function is generalized beyond the linear setting.  To this end, we consider the following class of \textit{bilevel quadratic programs} (BQPs):
\begin{subequations}
	\label{BQCP}
	\begin{align}
		[\textbf{BQP}]: \quad
		\min_{\mathbf{x}} \; & z_l(\mathbf{x}, \mathbf{y}^*)
		\label{obj: leader quadratic} \\
		\text{s.t. } 
		& \mathbf{x} \in X(\mathbf{y}^*) \label{cons: leader quadratic} \\
		& \mathbf{y}^* \in 
		\argmin_{\;\mathbf{y} \in Y(\mathbf{x})} \, z_f(\mathbf{y}). \label{cons: follower quadratic}
	\end{align}
\end{subequations}
Here, the decision-makers' objective functions are given by:
\begin{subequations}
	\begin{align}
		& z_l(\mathbf{x}, \mathbf{y}^*) :=	\tfrac{1}{2} \, \mathbf{x}^\top \mathbf{P}_l \, \mathbf{x} + \mathbf{c}_{l}^{\top}\mathbf{x} +\tfrac{1}{2}\, \mathbf{y}^{*\top} \mathbf{Q}_l \, \mathbf{y}^* + 
		\mathbf{d}_{l}^{\top}\mathbf{y}^*, \label{eq: leader's objective} \\
		& z_f(\mathbf{y}) := \tfrac{1}{2} \mathbf{y}^\top \mathbf{Q}_f \mathbf{y} + \mathbf{d}_{f}^{\top}\mathbf{y}, \label{eq: follower's objective}
	\end{align}
\end{subequations}
where $\mathbf{P}_l \in \mathbb{Q}^{n_l \times n_l}$,
$\mathbf{Q}_l \in \mathbb{Q}^{n_f \times n_f}$ and $\mathbf{Q}_f \in \mathbb{Q}^{n_f \times n_f}$. 

To the best of our knowledge, the computational complexity of general bilevel quadratic programs in fixed dimensions has not been systematically studied. Instead, the existing literature has primarily focused on algorithmic approaches for BQPs with a convex quadratic lower-level problem. These include, for example, KKT-based reformulations and branch-and-bound schemes~\cite{Bard1988,Bard1990}, descent methods~\cite{Vicente1994}, and sequential quadratic or smoothing techniques~\cite{Etoa2010,Etoa2011}. 

\begin{table}[h!]
	\centering \small
	\setstretch{1.25}
	\renewcommand{\arraystretch}{1.2}
		\scalebox{0.95}{\begin{tabular}{@{}llll@{}}
		\toprule
		& Fixed parameter & Optimistic version & Pessimistic version \\
		\midrule
		\multirow{2}{*}{Follower (linear)} 
		& $n_f$ 
		& $\mathrm{P}$~\cite{Deng1998,Liu1995} 
		& $\boldsymbol{\mathrm{NP}}$\textbf{-complete } \\[2pt]
		
		& $m_f$ 
		& $\boldsymbol{\mathrm{P}}$ 
		& $\boldsymbol{\mathrm{P}}$ \\[6pt]
		
		\midrule
		
		\multirow{2}{*}{Follower (convex quadratic)} 
		& $n_f$ 
		& $\boldsymbol{\mathrm{P}}$ (Theorem \ref{theorem 4})
		& $\mathrm{NP}$-hard$^{*}$ \\[2pt]
		
		& $m_f$ 
		& \textit{Open} 
		& $\boldsymbol{\mathrm{NP}}$\textbf{-hard}$^{*}$ (Theorem \ref{theorem 5}) \\[2pt]
		
		\midrule
		
		\multirow{2}{*}{Follower (nonconvex quadratic)} 
		& $n_f$
		& \multirow{2}{*}{$\left.
			\vcenter{\hbox{\rule{0pt}{2.7em}}}
			\right\}
			\ \boldsymbol{\mathrm{NP}}$\textbf{-hard}$^{*}$ (Theorem \ref{theorem 6})}
		& $\mathrm{NP}$-hard$^{*}$ \\[2pt]
		
		& $m_f$
		&
		& $\mathrm{NP}$-hard$^{*}$ \\[2pt]		
		\bottomrule
	\end{tabular}}
	\caption{
		Computational complexity of bilevel linear and quadratic programs, [\textbf{BLP}] and [\textbf{BQP}], in fixed dimensions.
		Theorems \ref{theorem 5} and \ref{theorem 6} are established under a \textit{linear} leader objective, whereas Theorem~\ref{theorem 4} assumes a \textit{convex quadratic} leader objective.   An asterisk indicates that the corresponding complexity result is a lower bound.  Bold entries indicate our results, and the remaining $\mathrm{NP}$-hard entries follow by containment of $\mathrm{NP}$-hard subclasses.
	}
	\label{tab: complexity 2}
\end{table}

 In essence, we demonstrate that the form of the follower's objective function in BQPs provides a source of computational hardness similar to that of coupling constraints in BLPs (see Table \ref{tab: complexity 2}). Our main results can be summarized as follows: 
\begin{itemize}
\item We show that the optimistic version of [\textbf{BQP}] with a fixed number of follower variables $n_f$ remains polynomially solvable when both the leader's and the follower's objective functions are convex quadratic. 
\item We demonstrate that the pessimistic version of~[\textbf{BQP}] with a fixed number of follower constraints~$m_f$ becomes $\mathrm{NP}$-hard, even when the leader's and the follower's objective functions are linear and convex quadratic, respectively. This result is established by extending the construction of Disser and Skutella \cite{Disser2018}  and requires no coupling constraints.  
\item Finally, we establish that the optimistic version of [\textbf{BQP}] becomes $\mathrm{NP}$-hard even when both $n_f$ and $m_f$ are fixed, provided that the follower's objective function is allowed to be nonconvex.
\end{itemize}

Unless $\mathrm{P}=\mathrm{NP}$, the results in Table~\ref{tab: complexity 2} reveal   underlying   complexity  transitions  between bilevel linear and quadratic optimization problems in fixed dimensions. In particular, replacing the follower's linear objective function with a convex quadratic one renders the corresponding pessimistic problem $\mathrm{NP}$-hard.  This transition has no counterpart in single-level optimization, where both linear and convex quadratic programs are polynomially solvable. Moreover, in contrast to the linear case, $\mathrm{NP}$-hardness for the convex quadratic case is established without using pessimistic coupling constraints. Second,  we demonstrate that replacing a convex quadratic follower objective with a nonconvex quadratic one  makes the corresponding optimistic problem $\mathrm{NP}$-hard.  Notably, the latter hardness result holds even when both follower dimensions are fixed  and small. 

The $\mathrm{NP}$-hardness results in Table~\ref{tab: complexity 2} provide lower bounds only; matching upper complexity bounds are not established.  Deriving  such bounds would require characterizing the complexity of verifying global lower-level optimality, particularly in the presence of a nonconvex quadratic follower objective, and lies beyond the scope of this note. In addition, the complexity of the optimistic problem with fixed~$m_f$ and a convex quadratic follower objective does not appear to follow directly from the techniques used in Theorems~\ref{theorem 4} and~\ref{theorem 5}, and remains open.


The remainder of the note is organized as follows. 
Section~\ref{subsec: blp model} introduces our assumptions and presents formulations of the optimistic and pessimistic versions of [\textbf{BLP}]. Sections~\ref{subsec: blp poly} and~\ref{subsec: blp np} establish polynomial-time solvability and $\mathrm{NP}$-hardness results for [\textbf{BLP}], respectively. The corresponding results for [\textbf{BQP}] are presented in Sections~\ref{subsec: bqp poly} and~\ref{subsec: bqp np}. Finally, Section~\ref{sec: conclusion} concludes the note and outlines directions for future research.

\textbf{Notation.} We use $\mathbb{R}_+$, $\mathbb{Z}_+$, and $\mathbb{Q}$ to denote the sets of nonnegative real numbers, nonnegative integers, and rational numbers, respectively. 
Vectors and matrices are denoted by boldface letters, with $\mathbf{e}_i$ and $\mathbf{1}$ representing the $i$-th unit and the all-ones~vectors of an appropriate dimension. 
Capital and calligraphic letters are used, respectively, for sets and collections of sets. For a polytope~$P$, $\mathrm{ext}(P)$ denotes its set of extreme points.

\section{Complexity of Bilevel Linear Programs} \label{sec: blp}
\subsection{Problem Definition and Value-Function Reformulation}
\label{subsec: blp model}

 We first state a standing assumption used throughout the note. Let the leader's feasible set~(\ref{eq: leader's feasible set}) be expressed as
\begin{equation} \label{eq: leader's feasible set 2}
	X(\mathbf{y}^*) =
	\big\{\mathbf{x}\in \tilde X :
	\mathbf{a}_l^{(j)\top}\mathbf{x}
	+
	\mathbf{g}_l^{(j)\top}\mathbf{y}^*
	\le h_l^{(j)},\ j\in J
	\big\},
\end{equation}
where
$\tilde X :=
\{\mathbf{x}\in\mathbb{R}^{n_l}_+:
\tilde{\mathbf A}_l\mathbf{x}\le \tilde{\mathbf h}_l\}$
is defined by the constraints in~(\ref{eq: leader's feasible set})
that do not depend on~$\mathbf{y}^*$. 
In line with standard conventions in the bilevel linear programming literature \cite{Hansen1992, Wiesemann2013}, we make the following assumption:
\begin{itemize}
	\item[\textbf{A1.}] 
	The feasible region $\tilde{X}$ is nonempty and bounded. Moreover, for
	every $\mathbf{x}\in \tilde{X}$, the follower's feasible set $Y(\mathbf{x})$ is
	nonempty and bounded. 
\end{itemize}

Assumption~\textbf{A1} ensures that the follower's optimization problem in (\ref{cons: follower}) has a nonempty and bounded feasible set and attains an optimal solution for every $\mathbf{x} \in \tilde{X}$. Importantly,  we do not treat the verification of Assumption~\textbf{A1} as part of the computational problem, and therefore all complexity statements are made for instances satisfying~\textbf{A1}. We also refer to \cite{Rodrigues2026} for the complexity results concerning boundedness in bilevel linear optimization.  

 Next, we consider the value-function reformulation of [\textbf{BLP}]; see, e.g., \cite{Dempe2002, Outrata1990}. Let the lower-level \textit{value function} $\varphi(\mathbf{x})$ represent the follower’s optimal objective function value in~[\textbf{BLP}] for a given feasible leader decision~ $\mathbf{x} \in \tilde{X}$. That is, under Assumption \textbf{A1}, 
\begin{equation} \label{eq: value function}
	\varphi(\mathbf{x})	:= \min_\mathbf{y} \, \Big\{
	\mathbf{d}_{f}^{\top}\mathbf{y}: \; \mathbf{y} \in Y(\mathbf{x}) \Big\} = \max_{\boldsymbol{\lambda} \geq 0} \Big\{
	(\mathbf{A}_f\mathbf{x} - \mathbf{h}_{f})^{\top}\boldsymbol{\lambda}
	:\;
	-\mathbf{G}_f^{\top}\boldsymbol{\lambda} \leq \mathbf{d}_{f}
	\Big\},
\end{equation}
where the last equality leverages strong duality for linear programs. 
By definition,~$\varphi(\mathbf{x})$ is a pointwise maximum of affine functions and is therefore piecewise linear and convex in~$\mathbf{x}$. Also, let 
\begin{equation} \label{eq: reaction set}
	R(\mathbf{x}) := \Big\{\mathbf{y} \in Y(\mathbf{x}): \; \mathbf{d}_{f}^{\top}\mathbf{y} = \varphi(\mathbf{x}) \Big\} = \Big\{\mathbf{y} \in Y(\mathbf{x}): \; \mathbf{d}_{f}^{\top}\mathbf{y} \leq \varphi(\mathbf{x}) \Big\}
\end{equation}
denote the follower’s optimal solution set, or \textit{reaction set}.
By the definition of $\varphi(\mathbf{x})$, the two representations of $R(\mathbf{x})$ using the equality “$=$” and the inequality “$\le$” are equivalent. Moreover, for any feasible $\mathbf{x} \in \tilde{X}$, the reaction set $R(\mathbf{x})$ is nonempty under Assumption~\textbf{A1}. 

Next, based on the definitions in (\ref{eq: value function}) and (\ref{eq: reaction set}), we introduce the optimistic and the pessimistic versions of~[\textbf{BLP}]. First, the optimistic problem can be expressed as
\begin{subequations}
	\label{OBLP}
	\begin{align}
		[\textbf{OBLP}]: \quad
		\min_{\mathbf{x}, \mathbf{y}^*} & \; 
		\mathbf{c}_{l}^{\top}\mathbf{x} + 
		\mathbf{d}_{l}^{\top}\mathbf{y}^* \label{obj: optimistic} \\
		\text{s.t. } 
		& \mathbf{x} \in X(\mathbf{y}^*) \label{cons: optimistic coulpling} \\
		& \mathbf{y}^* \in R(\mathbf{x}). \label{cons: optimistic optimality}
	\end{align}
\end{subequations}
On the other hand, the pessimistic version of [\textbf{BLP}] reads as
\begin{subequations}
	\label{PBLP}
	\begin{align}
		[\textbf{PBLP}]: \quad
		\min_{\mathbf{x}} \; & \max_{\bar{\mathbf{y}} \in R(\mathbf{x})} \;
	\mathbf{c}_{l}^{\top}\mathbf{x} + \mathbf{d}_{l}^{\top}\bar{\mathbf{y}} \label{obj: pessimistic}
		\\
		\text{s.t. } & 
		\mathbf{x} \in X(\mathbf{y}^*) \quad \forall \, \mathbf{y}^* \in R(\mathbf{x}). \label{cons: pessimistic coupling} 
	\end{align}
\end{subequations}

Notably, by using a standard epigraph reformulation, the uncertainty with respect to $\bar{\mathbf{y}}$ in the objective function (\ref{obj: pessimistic}) can be shifted to the coupling constraints (\ref{cons: pessimistic coupling}); see, e.g., \cite{Wiesemann2013}. 
Hence, in our complexity analysis  of [\textbf{PBLP}], 
we assume without loss of generality that the upper-level objective function is independent of~$\bar{\mathbf{y}}$, i.e., $\mathbf{d}_l = \mathbf{0}$.  Furthermore, under Assumption~\textbf{A1}, every feasible instance of [\textbf{PBLP}] attains an optimal solution. Indeed, by standard parametric linear programming arguments, the follower's value function $\varphi(\mathbf{x})$  and, for every $j \in J$, the worst-case value functions associated with the pessimistic coupling constraints 
\[
\psi_{j}(\mathbf{x})
:=
\max_{\mathbf{y}}\Big\{	\mathbf{g}_l^{(j)\top}\mathbf{y}: \,\mathbf{y}\in R(\mathbf{x})\Big\} = \max_{\mathbf{y}}\Big\{
	\mathbf{g}_l^{(j)\top}\mathbf{y} :\,
\mathbf{y}\in Y(\mathbf{x}),\;
\mathbf{d}_f^{\top}\mathbf{y}\leq \varphi(\mathbf{x})
\Big\} 
\] 
are continuous on the compact set $\widetilde X$; see, e.g., \cite{Gal1995}. Hence, the feasible set of [\textbf{PBLP}] is closed, and the minimum is attained. 

\subsection{Polynomial-Time Solvability} \label{subsec: blp poly}
In this section, we analyze the optimistic and the pessimistic formulations, [\textbf{OBLP}] and [\textbf{PBLP}], assuming that either the number of follower variables or constraints in (\ref{eq: follower's feasible set}) is fixed. The following result provides a concise proof of the existing results in \cite{Deng1998, Liu1995} and extends them to the case of a fixed number of follower~constraints.

\begin{theorem} \label{theorem 1}
 Under Assumption \textbf{A1}, the optimistic problem \upshape [\textbf{OBLP}] \itshape with a fixed number of either follower constraints $m_f$ or follower variables $n_f$ is polynomially solvable.
	\begin{proof}
		We begin with analyzing the dual feasible set in (\ref{eq: value function}) in two special cases, where either $m_f$ or~$n_f$ is fixed. First, when $m_f$ is fixed, the dual feasible set 
		\begin{equation} \label{dual feasible set}
			\Lambda := \Big\{\boldsymbol{\lambda} \in \mathbb{R}^{m_f}_+: \; -\mathbf{G}_f^{\top}\boldsymbol{\lambda} \leq \mathbf{d}_{f} \Big\}	
		\end{equation}
		has fixed dimension $m_f$. By definition, $\Lambda$ is defined by $n_f + m_f$ linear inequalities. Hence, extreme points, $\text{ext}(\Lambda)$, can be enumerated in polynomial time, e.g., by basis enumeration~over \[\binom{n_f + m_f}{m_f} = \tfrac{(n_f + m_f) \ldots (n_f + 1)}{m_f!} = O(n_f ^{m_f})\]
		active sets of constraints. That is, the number of extreme points, $|\text{ext}(\Lambda)|$, is bounded by~$O(n_f^{m_f})$. 
		
		Next, assume that $n_f$ is fixed. 
		Analogously, the number of extreme points in (\ref{dual feasible set}) is bounded from above~by
		\[\binom{n_f + m_f}{m_f} = \tfrac{(m_f + n_f) \ldots (m_f + 1)}{n_f!} = O(m_f ^{n_f}).\]
		Thus, in both cases, the value function (\ref{eq: value function}) can be expressed as
		\begin{equation} \label{eq: value function polynomial}
			\varphi(\mathbf{x}) = \max_{\boldsymbol{\lambda} \in \text{ext}(\Lambda)} \Big\{ (\mathbf{A}_f\mathbf{x} - \mathbf{h}_{f})^{\top}\boldsymbol{\lambda} \Big\} = \max_{i \in I} \Big\{ (\mathbf{A}_f\mathbf{x} - \mathbf{h}_{f})^{\top}\boldsymbol{\lambda}^{(i)} \Big\},
		\end{equation}
		where $|I|$ is polynomial in the size of the problem. Consequently, [\textbf{OBLP}] reads as
		\begin{subequations} \label{eq: optimistic polynomial}
			\begin{align}
				\min_{\mathbf{x}, \mathbf{y}^*, \mathbf{z}} & \;
				\mathbf{c}_{l}^{\top}\mathbf{x} + 
				\mathbf{d}_{l}^{\top}\mathbf{y}^* 
			 \\
				\text{s.t. } 
				& \mathbf{x} \in X(\mathbf{y}^*) \\
				& \mathbf{y}^* \in Y(\mathbf{x}) \\
				& \mathbf{d}_{f}^{\top}\mathbf{y}^* \leq \sum_{i \in I} (\mathbf{A}_f\mathbf{x} - \mathbf{h}_{f})^{\top}\boldsymbol{\lambda}^{(i)} z_i \\
				& \sum_{i \in I} z_i = 1 \\
				& \mathbf{z} \in \{0, 1\}^{|I|},
			\end{align}
		\end{subequations}
		where $z_i \in \{0,1\}$ indicates whether the maximum in (\ref{eq: value function polynomial}) is attained at the $i$-th term. As a result,~(\ref{eq: optimistic polynomial})~can~be tackled by solving $|I|$ linear programming problems with $z_i = 1$, $i \in I$, and selecting a solution with the smallest optimal objective function value. This observation concludes the proof. 
	\end{proof}	
\end{theorem}	
\begin{corollary} \label{corollary 1}
	Suppose that the leader's feasible set $X(\mathbf{y^*})$ is defined by \upshape (\ref{eq: leader's feasible set 2})  \itshape 
    and Assumption~\textbf{A1} holds. Then, ~\upshape[\textbf{PBLP}] \itshape with either a fixed number of follower constraints $m_f$ or a fixed number of variables~$n_f$, and a fixed number of coupling constraints $|J|  \geq 1$, is polynomially~solvable.
	\begin{proof}
		The result is implied by Theorem 3.3 in \cite{Henke2026}. That is, [\textbf{PBLP}] with $|J|$ coupling constraints can be equivalently reformulated as [\textbf{OBLP}] with coupling constraints, where the follower has $(m_f + 1) |J|$ constraints and $n_f |J|$ variables. This observation along with Theorem \ref{theorem 1} concludes the~proof. 
	\end{proof}
\end{corollary}

Alternatively, Corollary \ref{corollary 1} can be derived by applying a dual reformulation to the pessimistic coupling constraints (\ref{cons: pessimistic coupling}). In particular, as in the proof of Theorem \ref{theorem 1}, each resulting dual constraint can be expressed as a finite union of linear inequalities and can therefore be handled using a polynomial number of binary variables. However, the total number of feasible assignments of these binary variables is polynomial, and thus amenable to total enumeration, only when the number of coupling constraints~$|J|$~is~fixed.

Based on the above intuition, the computational complexity of [\textbf{PBLP}] with a fixed number of follower variables or follower constraints, but a nonfixed number of coupling constraints, remains unresolved. Indeed, in Section \ref{subsec: blp np} we demonstrate that [\textbf{PBLP}] with fixed $n_f$ is strongly $\mathrm{NP}$-hard. 
On a positive note, by leveraging some fundamental results from computational geometry, we demonstrate that [\textbf{PBLP}] with fixed $m_f$ remains polynomially solvable, even when the number of coupling constraints is not fixed; see Table~\ref{tab:complexity}. The following results~hold. 
 
\begin{theorem} \label{theorem 2}
	  Under Assumption~\textbf{A1}, \upshape[\textbf{PBLP}] \itshape with a fixed number of follower constraints $m_f$ is polynomially solvable.
	\begin{proof}
		
 First, we recall from Section \ref{subsec: blp model} that without loss of generality $\mathbf{d}_l = \mathbf{0}$. Then, using the definition of the leader's feasible set (\ref{eq: leader's feasible set 2}), the pessimistic problem [\textbf{PBLP}] reads as
		\begin{align*}
			\min_{\mathbf{x}} \; & 
			\mathbf{c}_{l}^{\top} \mathbf{x}
			\\
			\text{s.t. } & \mathbf{x} \in \tilde{X} \\
			& \max_{\mathbf{y}^* \in R(\mathbf{x})} \mathbf{g}^{(j)\top}_l \mathbf{y}^* \leq h^{(j)}_l - \mathbf{a}^{(j)\top}_l \mathbf{x}, \; j \in J. \label{cons: pessimistic robust}
		\end{align*}
		

		To proceed, we introduce a new variable $\mathbf{w} \in \mathbb{R}^{m_f}$ such that $\mathbf{w} := \mathbf{A}_f \mathbf{x}$, and the dimension of~$\mathbf{w}$ is fixed by our assumption. Also, we redefine
		\[\tilde{Y}(\mathbf{w}) := \Big\{\mathbf{y} \in \mathbb{R}_+^{n_f}: \, \mathbf{w} + \mathbf{G}_f \mathbf{y} \leq \mathbf{h}_f \Big\} \, \; \text{ and } \; \, \tilde{\varphi}(\mathbf{w}) := \min_\mathbf{y} \, \Big\{
		\mathbf{d}_{f}^{\top}\mathbf{y}: \; \mathbf{y} \in \tilde{Y}(\mathbf{w}) \Big\}.\]
	 Then, [\textbf{PBLP}] can be expressed in the following form:
		\begin{subequations} \label{pessimistic robust 1}
			\begin{align}
				\min_{\mathbf{x}, \mathbf{w}, t} \; & 
				\mathbf{c}_{l}^{\top} \mathbf{x}
				\\
				\text{s.t. } & \mathbf{x} \in \tilde{X} \label{cons: pessimistic robust 1 1} \\
				& \mathbf{w} = \mathbf{A}_f \mathbf{x} \label{cons: pessimistic robust 1 2}\\ 
				& t = \tilde{\varphi}(\mathbf{w}) \label{cons: pessimistic robust 1 3} \\
				& \max_{\mathbf{y}^* \in \tilde{R}(\mathbf{w}, t)} \mathbf{g}^{(j)\top}_l \mathbf{y}^* \leq h^{(j)}_l - \mathbf{a}^{(j)\top}_l \mathbf{x}, \; j \in J, \label{cons: pessimistic robust 2}
			\end{align}
		\end{subequations}
		where $\tilde{R}(\mathbf{w}, t) := \{\mathbf{y} \in \tilde{Y}(\mathbf{w}): \; \mathbf{d}_{f}^{\top}\mathbf{y} = t \}$. Notably, each coupling constraint in~(\ref{cons: pessimistic robust 2}) is linked to the same decision-dependent set $\tilde{R}(\mathbf{w}, t)$.
		
		For each $j \in J$, we consider the parametric linear program in the left-hand side of (\ref{cons: pessimistic robust 2}), i.e.,
		\begin{equation} \nonumber \label{parametric LP pessimistic}
		\max_{\mathbf{y}^* \in \tilde{R}(\mathbf{w}, t)} \mathbf{g}^{(j)\top}_l \mathbf{y}^* = \max_{\mathbf{y}^* \geq \mathbf{0}} \Big\{\mathbf{g}^{(j)\top}_l \mathbf{y}^*: \; \mathbf{G}_f \mathbf{y}^* \leq \mathbf{h}_f - \mathbf{w}, \; \mathbf{d}_{f}^{\top}\mathbf{y}^* = t\Big\}. 
		\end{equation}
		 By Assumption \textbf{A1}, for any feasible solution $(\mathbf{w}, t)$ of (\ref{pessimistic robust 1}), $\tilde{R}(\mathbf{w}, t)$ is a nonempty and bounded polyhedron, and therefore it has at least one extreme point. Moreover, any extreme point $\mathbf{v} \in \text{ext}(\tilde{R}(\mathbf{w}, t))$ is determined by $n_f$ linearly independent active constraints, including the equality constraint~$\mathbf{d}_f^{\top}\mathbf{v}=~t$ and $n_f - 1$ out of 
		$n_f + m_f$ inequality constraints $\mathbf{v} \geq 0$ and
		$\mathbf{G}_f \mathbf{v} \leq \mathbf{h}_f - \mathbf{w}$. In the following, we assume that $\mathbf{d}_f \neq \mathbf{0}$, so that the equality constraint is nonredundant; the case $\mathbf{d}_f = \mathbf{0}$ can be treated~analogously. 
		
		 Let $U := \{0, 1, \ldots, n_f + m_f\}$ denote the index set of all constraints in $\tilde{R}(\mathbf{w}, t)$,
		where index $0$ corresponds to the equality constraint. Then, we denote by $B \subseteq U$, with $0 \in B$ and $|B| = n_f$, the index set of the $n_f$ active constraints. Also, let $N := U \setminus B$ be the index set of the remaining~$m_f + 1$ inequality constraints. 
		
		Stacking the $n_f$ active constraints indexed by $B$ in matrix form~gives
		$\mathbf{M}_B \mathbf{v} = \mathbf{r}_B(\mathbf{w},t)$,
		where $\mathbf{M}_B \in \mathbb{R}^{n_f \times n_f}$ is nonsingular, and $\mathbf{r}_B(\mathbf{w},t) \in \mathbb{R}^{n_f}$ is affine in $(\mathbf{w},t)$.
		Moreover, 
		\begin{equation} \label{cons: vertex active}
		\mathbf{v}_B(\mathbf{w},t)
		:= \mathbf{M}_B^{-1} \, \mathbf{r}_B(\mathbf{w},t)
		\end{equation}
		is an extreme point of $\tilde{R}(\mathbf{w},t)$ if and only if it also satisfies the remaining nonactive constraints,~i.e.,
		\begin{equation} \label{cons: vertex non-active}
		\mathbf{M}_N \, \mathbf{v}_B(\mathbf{w},t) = \mathbf{M}_N \, \mathbf{M}_B^{-1} \, \mathbf{r}_B(\mathbf{w},t) \leq \mathbf{r}_N(\mathbf{w},t).
		\end{equation}
		Thus, for a fixed basis $B$, the extreme point $\mathbf{v}_B(\mathbf{w},t)$
		defined by equation (\ref{cons: vertex active}) is feasible whenever it satisfies $m_f + 1$ inequality constraints (\ref{cons: vertex non-active}). 
		
		Let \[\mathcal{B} = \Big\{B \subseteq U: \; 0 \in B, \; |B| = n_f \; \mbox{ and } \; \det(\mathbf{M}_B)\neq 0 \Big\}\]
		be the collection of all candidate bases. We observe that \[|\mathcal{B}| \leq \binom{n_f + m_f}{n_f - 1} = O(n_f^{m_f + 1}),\] when $m_f$ is fixed. Moreover, let
		\begin{equation} \nonumber
			W_B = \Big\{(\mathbf{w}, t) \in \mathbb{R}^{m_f + 1}: \text{ (\ref{cons: vertex non-active}) holds} 
			\Big\}
		\end{equation}
		be a feasible set for a basis $B \in \mathcal{B}$. 
		As a result, (\ref{pessimistic robust 1}) can be reformulated as
		\begin{subequations} \label{pessimistic robust 2}
			\begin{align}
				\min_{\mathbf{x}, \mathbf{w}, t} \; & 
				\mathbf{c}_{l}^{\top} \mathbf{x}
				\\
				\text{s.t. } & \text{(\ref{cons: pessimistic robust 1 1})--(\ref{cons: pessimistic robust 1 3}),} \\
				& \max_{B \in \mathcal{B}} \Big\{\mathbf{g}^{(j)\top}_l \mathbf{v}_B(\mathbf{w}, t): \; (\mathbf{w}, t) \in W_B \Big\} \leq h^{(j)}_l - \mathbf{a}^{(j)\top}_l \mathbf{x}, \; j \in J. \label{cons: pessimistic robust 3}
			\end{align}
		\end{subequations}
		
		Each set $W_B$ is defined by $m_f + 1$ linear inequalities with respect to $(\mathbf{w}, t) \in \mathbb{R}^{m_f + 1}$ and, hence, the collection of sets
		$\mathcal{W} := \{W_B: B \in \mathcal{B} \}$ 
		induces
		\[p = (m_f + 1)|\mathcal{B}| = O(n_f^{m_f + 1})\] linear inequalities in $\mathbb{R}^{m_f + 1}$. It is well known that an arrangement of $p$ hyperplanes in $\mathbb{R}^{m_f+1}$ induces at most $\tilde{p} = O(p^{m_f + 1})$ relatively open polyhedral cells $C_1, \ldots, C_{\tilde{p}}$, possibly of lower dimension, that partition $\mathbb{R}^{m_f + 1}$; see, e.g., \cite{Zaslavsky1975}. Notably,~$\tilde{p}$ is polynomial~in~$n_f$ when $m_f$ is fixed, and the cells can be constructed explicitly in polynomial time and space, namely $O(\tilde{p})$, by the Edelsbrunner–Seidel incremental construction~method~\cite{Edelsbrunner1986}.
		
		 Next, on each relatively open cell $C_k$, $k\in\{1,\ldots,\tilde p\}$, every inequality in $\mathcal W$ has a constant sign, i.e., it is either strictly satisfied, violated, or tight. Since the number of candidate bases~$|\mathcal{B}|$ is polynomial for fixed $m_f$, one can efficiently determine the subcollection of bases $ \mathcal{B}_{k} \subseteq \mathcal{B}$ that are feasible~on~$C_k$.
		We conclude that (\ref{pessimistic robust 2}) can be reduced to $\tilde{p}$ subproblems, indexed by $k \in \{1, \ldots, \tilde{p}\}$, of the form: 
		\begin{subequations} \label{pessimistic subproblem}
			\begin{align}
				\min_{\mathbf{x}, \mathbf{w}, t} \; & 
				\mathbf{c}_{l}^{\top} \mathbf{x}
				\\
				\text{s.t. } & \text{(\ref{cons: pessimistic robust 1 1})--(\ref{cons: pessimistic robust 1 3}),} \\
				& (\mathbf{w}, t) \in C_k \\
				& \mathbf{g}^{(j)\top}_l \mathbf{v}_B(\mathbf{w}, t) \leq h^{(j)}_l - \mathbf{a}^{(j)\top}_l \mathbf{x}, \quad \forall \, j \in J \; \mbox{ and } \; \forall \, B \in \mathcal{B}_{k}. \; \label{cons: pessimistic robust 4}
			\end{align}
		\end{subequations}

	 The analysis of subproblems \eqref{pessimistic subproblem} involves two technical difficulties. First, 
	 the equality constraint~(\ref{cons: pessimistic robust 1 3}) is nonlinear. In this regard, following the proof of Theorem \ref{theorem 1}, we have
		\[\tilde{\varphi}(\mathbf{w}) = \max_{\boldsymbol{\lambda} \in \text{ext}(\Lambda)} \Big\{ 	(\mathbf{w} - \mathbf{h}_{f})^{\top}\boldsymbol{\lambda} \Big\} = \max_{i \in I} \Big\{ (\mathbf{w} - \mathbf{h}_{f})^{\top}\boldsymbol{\lambda}^{(i)} \Big\},\]
		where $|I| = O(n_f^{m_f})$. Hence, $\tilde{\varphi}(\mathbf{w}) = t$ is equivalent to the following system of equations:
		\begin{equation} \nonumber
			\begin{cases}
				(\mathbf{w} - \mathbf{h}_{f})^{\top}\boldsymbol{\lambda}^{(i)} \leq t \quad \forall i \in I \\
				t \leq \sum_{i \in I} (\mathbf{w} - \mathbf{h}_{f})^{\top}\boldsymbol{\lambda}^{(i)} z_i \\
				\sum_{i \in I} z_i = 1 \\
				\mathbf{z} \in \{0, 1\}^{|I|}.
			\end{cases}
		\end{equation}
		Enumeration of $\mathbf{z}$ then reduces (\ref{pessimistic subproblem}) to $|I|$ linear programs of polynomial size. 

	Second, each relatively open cell $C_k$, $k \in \{1, \ldots, \tilde{p}\}$, may involve strict linear
	inequalities. To~address this, for each admissible vector
	$\mathbf z\in\{0,1\}^{|I|}$, we first solve the closure relaxation of~(\ref{pessimistic subproblem}), obtained by replacing~$C_k$ with
	its closure. 
 We then test whether the obtained optimal value is attained on~$C_k$ by fixing this value and maximizing a common slack in the strict inequalities defining $C_k$. If the optimal slack is positive, then the value is retained; otherwise, the value is attained only on the boundary of~$C_k$ and is therefore discarded, as the boundary is treated as a separate cell. As a result, each subproblem~(\ref{pessimistic subproblem}) reduces to a polynomial number of ordinary linear programs, followed by a polynomial-time verification procedure. 
 This observation concludes the~proof.
	\end{proof}
\end{theorem}
\begin{corollary} \label{corollary 2}
	  Under Assumption \textbf{A1},  the pessimistic problem \upshape [\textbf{PBLP}] \itshape with a fixed number of follower variables $n_f$ and a fixed number of leader variables $n_l$ is polynomially solvable.
	\begin{proof} The result follows from the proof of Theorem~\ref{theorem 2} by using fixed-dimensional~$\mathbf{x}$ directly as the parameter in the follower's problem~(\ref{cons: follower}), instead of introducing auxiliary variables~$\mathbf{w}=\mathbf{A}_f\mathbf{x}$. In particular, the corresponding hyperplane arrangement is constructed in the fixed-dimensional space~$\mathbb{R}^{n_l + 1}$, and the number of candidate bases $|\mathcal{B}|$ remains polynomial when $n_f$ is fixed. 
	\end{proof}
\end{corollary}

\subsection{\textit{NP}-Hardness Result} \label{subsec: blp np}
In this section, we demonstrate that the pessimistic problem [\textbf{PBLP}] with fixed $n_f$ and a nonfixed number of coupling constraints $|J|$ is $\mathrm{NP}$-hard in the strong sense. 
The proof leverages a polynomial-time reduction from the \textit{maximum independent set} (MIS) problem, which is known to be strongly~$\mathrm{NP}$-hard~\cite{Garey1979}. A decision version of the MIS problem is stated as~follows:
\begin{itemize}
	\item[] \textbf{[MIS-D]}: Given $q \in \mathbb{Z}_+$ and a graph $G = (V, E)$, 
	where $V$ is a set of vertices and $E$ is a set of edges in $G$, does there exist a subset $S \subseteq V$ such that no two vertices in $S$ are adjacent, and~$|S| \geq q$?
\end{itemize}

Let 
$V = \{v_k: k \in K\}$, where $K = \{1, \ldots, |V|\}$. 
 For each $k \in K$, we define a binary variable $x_k \in \{0, 1\}$ that equals $1$ if the vertex $v_k$ is included in~$S$, and $0$ otherwise. Then, the MIS problem can be viewed as the following integer linear program \cite{Bomze1999}: 
\begin{subequations} \label{maximum independent set}
	\begin{align}
		\mathrm{OPT}(G) := \max_{\mathbf{x}} & \sum_{k\in K} x_k \\
		\mbox{s.t. } &
		x_i+x_j \leq 1 \quad \forall \, (v_i, v_j) \in E \\
		& x_k \in \{0,1\} \quad \forall k \in K.
	\end{align}
\end{subequations}
In particular, the answer to [\textbf{MIS-D}] is ``yes'' if and only if $\mathrm{OPT}(G) \geq q$. The following lemma, which exploits the uniqueness of a discrete distribution determined by its first- and second-order moments, plays a central role in our reduction.

\begin{lemma} \label{lemma 1}
For each $k \in K$, $x_k \in \{0, 1\}$ holds if and only if $x_k \in [0, 1]$ and the following constraint is~satisfied:
\begin{subequations} \label{disjunctive constraint}
	\begin{align}
	 \min_{\boldsymbol{\lambda}, \bar{\boldsymbol{\lambda}}} & \Big\{\mathbf{x}^\top \boldsymbol{\lambda} + \bar{\mathbf{x}}^\top \bar{\boldsymbol{\lambda}} \Big\} 
		\leq 0 \\
		\mbox{\upshape s.t. } &	\boldsymbol{\lambda} \geq \mathbf{0}, \; \bar{\boldsymbol{\lambda}} \geq \mathbf{0} \\
		& \sum_{i \in K} \big(\lambda_i + \bar{\lambda}_i \big) = 1 \label{cons: disjunctive 1}\\
		& \sum_{i \in K} i\big(\lambda_i + \bar{\lambda}_i \big) = k \label{cons: disjunctive 2} \\
		& \sum_{i \in K} i^2\big(\lambda_i + \bar{\lambda}_i \big) = k^2, \label{cons: disjunctive 3}
	\end{align}
\end{subequations}
where $\boldsymbol{\lambda} \in \mathbb{R}_+^{|V|}$, $\bar{\boldsymbol{\lambda}} \in \mathbb{R}_+^{|V|}$ and $\bar{\mathbf{x}} = \mathbf{1} - \mathbf{x}$. 
\begin{proof}
	For each $i \in K$, let $\tau_i := \lambda_i + \bar{\lambda}_i \geq 0$. Then, the system of linear equalities (\ref{cons: disjunctive 1})--(\ref{cons: disjunctive 3}) can be expressed as
\begin{equation} \label{linear system}
	\left\{
	\begin{aligned}
		& \sum_{i \in K} \tau_i = 1\\
		&\sum_{i \in K} i\,\tau_i = k\\
		& \sum_{i \in K} i^2 \tau_i = k^2.
	\end{aligned}
	\right.
\end{equation} 
Notably,
\[\sum_{i \in K} (i - k)^2 \tau_i = \sum_{i \in K} (i^2 - 2i k + k^2) \tau_i = \sum_{i \in K} i^2 \tau_i - 2k \sum_{i \in K} i \tau_i + k^2 \sum_{i \in K} \tau_i = k^2 -2k^2 + k^2 = 0.\]
Since $(i-k)^2 \geq 0$ and $\tau_i \geq 0$ for all $i \in K$, we observe that $\tau_i = 0$ for all $i \neq k$.
Combined with the constraint $\sum_{i\in K}\tau_i = 1$, it implies that $\tau_k = 1$.
Thus, the system of equations (\ref{linear system}) has a \textit{unique} nonnegative solution $\boldsymbol{\tau}^*$ such that $\tau^*_i = 1$ if $i = k$, and $\tau^*_i = 0$ otherwise.

As a result, for each $k \in K$, (\ref{disjunctive constraint}) reads as
\begin{equation} \label{disjunctive constraint 2}
		\min_{\boldsymbol{\lambda} \geq \mathbf{0}, \, \bar{\boldsymbol{\lambda}} \geq \mathbf{0} } \Big\{\mathbf{x}^\top \boldsymbol{\lambda} + \bar{\mathbf{x}}^\top \bar{\boldsymbol{\lambda}} : \; \lambda_k + \bar{\lambda}_k = 1, \; \lambda_i + \bar{\lambda}_i = 0 \quad \forall i \in K \setminus \{k\} \Big\} 
		\leq 0. 
	\end{equation} 
In particular, (\ref{disjunctive constraint 2}) implies that
\[\min_{\lambda_k \geq 0, \; \bar{\lambda}_k \geq 0 } \Big\{x_k \lambda_k + \bar{x}_k \bar{\lambda}_k: \; \lambda_k + \bar{\lambda}_k = 1 \Big\} \leq 0,\]
and therefore $\min \{x_k, \bar{x}_k\} = \min \{x_k, 1 - x_k\} \leq 0$.
We conclude that (\ref{disjunctive constraint}) with $x_k \in [0, 1]$ implies that $x_k \in \{0, 1\}$. Conversely, if $x_k \in \{0,1\}$, then $\min\{x_k,1-x_k\} = 0 \leq 0$ and, hence, the set of constraints~(\ref{disjunctive constraint}) is satisfied. 
\end{proof}	
\end{lemma} 
\begin{theorem} \label{theorem 3}
	 Under Assumption \textbf{A1}, the pessimistic problem \upshape [\textbf{PBLP}] \itshape with a fixed number of follower variables $n_f$ is strongly $\mathrm{NP}$-complete.
	\begin{proof}
		
	    The decision version
	    of [\textbf{PBLP}] belongs to~$\mathrm{NP}$ due to Theorem 3 in~\cite{Buchheim2023}. In particular, the same certificate argument can be applied to each pessimistic coupling constraint.

		To establish strong $\mathrm{NP}$-hardness, we  demonstrate that the MIS problem (\ref{maximum independent set}) reduces to an instance of [\textbf{PBLP}] with a fixed number of follower variables $n_f$. In particular, by Lemma \ref{lemma 1}, (\ref{maximum independent set})~can be expressed~as 
		\begin{subequations} \label{maximum independent set 2}
			\begin{align}
				\mathrm{OPT}(G) = \max_{\mathbf{x}} & \sum_{i\in K} x_i \\
				\mbox{s.t. } &
				x_i+x_j \leq 1 \quad \forall \, (v_i, v_j) \in E \\
				& 0 \leq x_k \leq 1 \quad \forall \, k \in K \\
				& \min_{\boldsymbol{\lambda} \geq \mathbf{0}, \, \bar{\boldsymbol{\lambda}} \geq \mathbf{0}} \Big\{\mathbf{x}^\top \boldsymbol{\lambda} + (\mathbf{1} - \mathbf{x})^\top \bar{\boldsymbol{\lambda}}: \; (\ref{cons: disjunctive 1})-(\ref{cons: disjunctive 3})\Big\} 
				\leq 0 \quad \forall \, k \in K. \label{cons: disjunctive constraint}
			\end{align}
		\end{subequations}
	
	Let $y_i \in \mathbb{R}$, $i \in \{1, 2, 3\}$, be dual variables corresponding to the equality constraints (\ref{cons: disjunctive 1})--(\ref{cons: disjunctive 3}). 
	 By applying linear programming duality to (\ref{cons: disjunctive constraint}) for each $k \in K$, we obtain the following reformulation~of~(\ref{maximum independent set}): 
		\begin{subequations} \label{maximum independent set 3}
			\begin{align}
				\mathrm{OPT}(G) = \max_{\mathbf{x}} & \sum_{i\in K} x_i \\
				\mbox{s.t. } &
				x_i+x_j \leq 1 \quad \forall \, (v_i, v_j) \in E \label{cons: independent set final 1} \\
				& 0 \leq x_k \leq 1 \quad \forall \, k \in K \label{cons: independent set final 2}\\
				& \max_{\mathbf{y} \in Y(\mathbf{x})} \Big\{-y_1 - k y_2 - k^2 y_3 \Big\} 
				\leq 0 \quad \forall k \in K, \label{cons: independent set final 3}
			\end{align}
		\end{subequations}
		where the dual feasible set
		\begin{equation} \nonumber
			 Y(\mathbf{x}) = \Big\{\mathbf{y} \in \mathbb{R}^3: \; \mathbf{x} + y_1 \mathbf{1} + y_2 \mathbf{v} + y_3 \mathbf{v}^2 \geq \mathbf{0}, \; \mathbf{1} - \mathbf{x} + y_1 \mathbf{1} + y_2 \mathbf{v} + y_3 \mathbf{v}^2 \geq \mathbf{0} \Big\}
		\end{equation}
	does not depend on $k$, $\mathbf{v} := (1, 2,\ldots, |V|)^\top$ and \(\mathbf{v}^2\) denotes its componentwise square. In particular, strong duality holds since the primal problem in the left-hand side of (\ref{cons: disjunctive constraint}) is feasible and has a finite optimal value. 
	
	 As a result, (\ref{maximum independent set 3}) reduces to the following instance of [\textbf{PBLP}]:
	\begin{subequations} \label{maximum independent set 4}
		\begin{align}
			\mathrm{OPT}(G) = \min_{\mathbf{x}} & -\sum_{i\in K} x_i \\
			\mbox{s.t. } & \text{(\ref{cons: independent set final 1})--(\ref{cons: independent set final 2})}, \\ 
			& -y_1 - k y_2 - k^2 y_3 
			\leq 0 \quad \forall\, \mathbf{y} \in Y(\mathbf{x}),\quad \forall k \in K.
		\end{align}
	\end{subequations}	
	Here, \(\mathbf{d}_l = \mathbf{d}_f = \mathbf{0}\), and therefore the follower's reaction set \(R(\mathbf{x})\) coincides with  their  feasible set \(Y(\mathbf{x})\). Notably,
	 the follower variables $y_i \in \mathbb{R}$, $i \in \{1,2,3\}$, are unrestricted in sign, while $Y(\mathbf{x})$ is unbounded by construction. 
	To address these issues, we first replace $y_i \in \mathbb{R}$, $i \in \{1,2,3\}$, with two nonnegative variables $y_i^+ \in \mathbb{R}_+$ and $y_i^- \in \mathbb{R}_+$ such that $y_i = y_i^+ - y_i^-$. Second, it is rather straightforward to verify that, for each $k \in K$, the dual problem in (\ref{cons: independent set final 3}) admits an optimal solution
	\[(y^*_1,y^*_2,y^*_3)^\top=\beta_k(k^2-1,-2k,1)^\top,\]
	where $\beta_k:=\min\{x_k,1-x_k\} \in [0,1]$.
	Therefore, the bounds $y^+_i \leq |V|^2$ and $y^-_i \leq |V|^2$, $i\in\{1,2,3\}$, can be added to the definition of $Y(\mathbf{x})$ without affecting the reduction. We conclude that the modified construction satisfies Assumption \textbf{A1}. Finally, a ``yes'' instance of [\textbf{MIS-D}] implies a ``yes'' instance of the threshold version of (\ref{maximum independent set 4}), and vice versa. This observation implies the~result. 
	\end{proof}
\end{theorem}

Theorem~\ref{theorem 3} has two important implications. First, together with Theorem~\ref{theorem 1}, it demonstrates that, under comparable assumptions, pessimistic BLPs can be computationally harder than their optimistic counterparts. Second, in view of Corollary~\ref{corollary 1}, the  hardness  of pessimistic BLPs is tied to the presence of a nonfixed number of coupling constraints.

\section{Complexity of Bilevel Quadratic Programs} \label{sec: bqp}
In this section, we investigate whether the polynomial-time solvability results established in Theorems~\ref{theorem 1} and~\ref{theorem 2} extend to bilevel quadratic programs~[\textbf{BQP}]. In particular, we show that seemingly mild generalizations of the follower's objective function in (\ref{cons: follower}) can lead to fundamentally different computational complexity behavior of [\textbf{BQP}]; recall Table~\ref{tab: complexity 2}.

Let the leader's and the follower's objective functions, $z_l(\mathbf{x}, \mathbf{y}^*)$ and $z_f(\mathbf{y})$, be defined by equations~(\ref{eq: leader's objective}) and (\ref{eq: follower's objective}), respectively. Then, the optimistic version of~[\textbf{BQP}] is given by:
\begin{subequations}
	\label{OBQP}
	\begin{align}
		[\textbf{OBQP}]: \quad
		\min_{\mathbf{x}, \mathbf{y}^*} \; & z_l(\mathbf{x}, \mathbf{y}^*)
		\label{obj: obqp} \\
		\text{s.t. } 
		& \mathbf{x} \in X(\mathbf{y}^*) \label{cons: obqp leader} \\
		& \mathbf{y}^* \in 
		\argmin_{\;\mathbf{y} \in Y(\mathbf{x})} \, z_f(\mathbf{y}).
		\label{cons: obqp follower}
	\end{align}
\end{subequations}
Furthermore, the pessimistic version of [\textbf{BQP}] reads as
\begin{subequations}
	\label{PBQP}
	\begin{align}
		[\textbf{PBQP}]: \quad
		\min_{\mathbf{x}} \; &
		\max_{\bar{\mathbf{y}} \in \tilde{R}(\mathbf{x})} 
		z_l(\mathbf{x}, \bar{\mathbf{y}})
		\label{obj: pbqp} \\
		\text{s.t. } 
		& \mathbf{x} \in X(\mathbf{y}^*)
		\quad \forall \mathbf{y}^* \in \tilde{R}(\mathbf{x}),
		\label{cons: pbqp leader}
	\end{align}
\end{subequations}
where
$\tilde{R}(\mathbf{x})
:=
\argmin_{\;\mathbf{y} \in Y(\mathbf{x})} \, z_f(\mathbf{y})$
is the follower's reaction set for a given feasible leader decision~$\mathbf{x}$.
 Notably, Assumption~\textbf{A1} guarantees that $\tilde{R}(\mathbf{x})$ is nonempty and compact for every $\mathbf{x}\in\widetilde X$. 

\subsection{Polynomial-Time Solvability}
\label{subsec: bqp poly}
In this section, we assume that [\textbf{BQP}] is a bilevel convex quadratic program, i.e., the matrices~$\mathbf{P}_l$,~$\mathbf{Q}_l$, and $\mathbf{Q}_f$ are symmetric positive semidefinite. By combining Carathéodory's theorem with KKT optimality conditions, we show that, under the outlined convexity assumption,~[\textbf{OBQP}] with fixed $n_f$ remains polynomially solvable; recall Table \ref{tab: complexity 2}. The following result holds. 

\begin{theorem} \label{theorem 4}
Under Assumption \textbf{A1}, the optimistic problem \upshape [\textbf{OBQP}] \itshape with $\mathbf{P}_l \succeq \mathbf{0}$, $\mathbf{Q}_l \succeq \mathbf{0}$, $\mathbf{Q}_f \succeq \mathbf{0}$, and a fixed number of follower variables $n_f$ is polynomially solvable. 
\begin{proof}

Since $\mathbf{Q}_f \succeq \mathbf{0}$ and Assumption~\textbf{A1} holds, the KKT optimality conditions for the follower's problem in~(\ref{cons: follower quadratic}) are necessary and sufficient.
Applying these conditions yields the following single-level reformulation of [\textbf{OBQP}]:
\begin{subequations}
	\label{BQP KKT}
	\begin{align}
		\min_{\mathbf{x}, \mathbf{y}^*, \boldsymbol{\lambda}, \boldsymbol{\nu}} \; & z_l(\mathbf{x}, \mathbf{y}^*)\\
		\text{s.t. } 
		& \mathbf{x} \in X(\mathbf{y}^*) \label{cons: leader KKT quadratic} \\
		& \mathbf{y}^* \in Y(\mathbf{x}) \label{cons: follower KKT quadratic} \\
		& 	\boldsymbol{\lambda} \geq \mathbf{0}, \; \boldsymbol{\nu} \geq \mathbf{0} \label{cons: dual KKT quadratic} \\
		& \mathbf{Q}_f \mathbf{y}^* + \mathbf{d}_f + \mathbf{G}_f^\top \boldsymbol{\lambda} - \boldsymbol{\nu} = \mathbf{0} \label{cons: stationarity KKT quadratic} \\
		& \boldsymbol{\nu}^\top \mathbf{y}^* = 0 \label{cons: complementarity condition 1 quadratic} \\
		& \boldsymbol{\lambda}^\top (\mathbf{h}_f - \mathbf{A}_f \mathbf{x} - \mathbf{G}_f \mathbf{y}^* ) = 0 \label{cons: complementarity condition 2 quadratic} .
	\end{align}
\end{subequations}
Constraints (\ref{cons: stationarity KKT quadratic}) and (\ref{cons: complementarity condition 1 quadratic})--(\ref{cons: complementarity condition 2 quadratic}) ensure stationarity and complementary slackness, respectively. 

Let the follower's feasible set (\ref{eq: follower's feasible set}) be written as
\begin{equation} \label{eq: follower's feasible set 2}
	Y(\mathbf{x}) =
	\left\{
	\mathbf{y} \in \mathbb{R}^{n_f}_+ :
	\mathbf{a}_f^{(j)\top}\mathbf{x}
	+
	\mathbf{g}_f^{(j)\top}\mathbf{y}
	\le h_f^{(j)},
	\quad j \in \{1,\ldots,m_f\}
	\right\},
\end{equation}
and denote by
\[J(\mathbf{x},\mathbf{y}^*)
:=
\Big\{
j :
\mathbf a_f^{(j)\top}\mathbf x
+
\mathbf g_f^{(j)\top}\mathbf y^*
=
h_f^{(j)}
\Big\} \;
\text{ and }
\; I(\mathbf y^*)
:=
\Big\{i:y_i^*=0\Big\},\] respectively, the sets of active follower and nonnegativity constraints at $\mathbf{y} = \mathbf{y}^*$.
Notably, (\ref{cons: stationarity KKT quadratic}) and~(\ref{cons: complementarity condition 1 quadratic})--(\ref{cons: complementarity condition 2 quadratic}) imply that
the vector $-(\mathbf Q_f\mathbf y^*+\mathbf d_f)$
belongs to the cone generated by the active constraint normals, i.e.,
\[
-(\mathbf Q_f\mathbf y^*+\mathbf d_f)
= \sum_{j\in J(\mathbf x,\mathbf y^*)}
\lambda_j \mathbf g_f^{(j)}
-
\sum_{i\in I(\mathbf y^*)}
\nu_i \mathbf e_i.
\]

By conic Carathéodory's theorem, $-(\mathbf Q_f\mathbf y^*+\mathbf d_f) \in \mathbb{R}^{n_f}$ can be represented using at most~$n_f$ such normals; see, e.g., \cite{Rockafellar1997}.
Hence, it is sufficient to
enumerate all pairs of subsets
$J' \subseteq \{1,\ldots,m_f\}$ and~$I' \subseteq \{1,\ldots,n_f\}$, such that
$|J'|+|I'| \leq n_f$. 
The number of such pairs is given by: 
\[\sum_{i = 0}^{n_f} \binom{n_f + m_f}{i} = O\Big((n_f + m_f)^{n_f}\Big)\]
and is therefore polynomial for fixed $n_f$. Moreover, since $\mathbf{P}_l \succeq \mathbf{0}$ and $\mathbf{Q}_l \succeq \mathbf{0}$, for fixed~$J'$ and $I'$, problem (\ref{BQP KKT})~reduces to the following convex quadratic program with linear constraints:
\begin{subequations}
	\label{BQP KKT 2}
	\begin{align}
		\min_{\mathbf{x}, \mathbf{y}^*, \boldsymbol{\lambda}, \boldsymbol{\nu}} \; & z_l(\mathbf{x}, \mathbf{y}^*)\\
		\text{s.t. } 
		& \text{(\ref{cons: leader KKT quadratic})--(\ref{cons: dual KKT quadratic})}, \\
		& -(\mathbf Q_f\mathbf y^*+\mathbf d_f)
		= \sum_{j\in J'}
		\lambda_j \mathbf g_f^{(j)}
		-
		\sum_{i\in I'}
		\nu_i \mathbf e_i \\
 & \mathbf a_f^{(j)\top}\mathbf x + \mathbf g_f^{(j)\top}\mathbf y^* = h_f^{(j)} \quad \forall j \in J' \\ & y^*_i = 0 \quad \forall i \in I' \\ & \lambda_j = 0 \quad \forall j \notin J' \\ & \nu_i = 0 \quad \forall i \notin I'.
	\end{align}
\end{subequations}
Since convex quadratic programs can be solved in polynomial time, the result follows.
\end{proof}
\end{theorem}

\subsection{\textit{NP}-Hardness Results}
\label{subsec: bqp np}
First, we demonstrate that the pessimistic problem [\textbf{PBQP}] with fixed $m_f$ becomes $\mathrm{NP}$-hard, even when the follower objective is convex quadratic and the leader objective remains linear; recall Table \ref{tab: complexity 2}. Surprisingly, this result relies on the construction of Disser and Skutella~\cite{Disser2018}, which, as we demonstrate below, also establishes the $\mathrm{NP}$-hardness of \textit{optimistic} BLPs with a single leader decision variable ($n_l = 1$); recall the first row of Table \ref{tab:complexity}.

 Let $G=(V,E)$ be a directed graph with arc capacities $\mathbf u\in\mathbb Q^{|E|}_+$, per-unit flow costs $\mathbf c\in\mathbb Q^{|E|}$, and designated source and destination nodes $s,t\in V$. Then, the parametric min-cost flow problem for a given total flow $x \in [0, \bar{x}]$ is given by:
\begin{subequations} \label{min cost flow problem}
	\begin{align}
		\min_{\mathbf{y}} & \sum_{e \in E} c_e y_e \label{obj: min cost flow} \\
		\text{s.t. } &
		0 \leq y_e \leq u_e \quad \forall e \in E, \label{cons: min cost flow 1} \\
		& \sum_{e \in E^+(v)} y_e - \sum_{e \in E^-(v)} y_e =
		\begin{cases}
			x, \mbox{ if } v = s,\\
			-x, \mbox{ if } v = t,\\
			0, \mbox{ if } v \in V \setminus \{s, t\},
		\end{cases}
		\label{cons: min cost flow 2}
	\end{align}
\end{subequations}
where $E^+(v)$ and $E^-(v)$ are, respectively, the sets of arcs directed out of and into node $v \in V$. The following intermediate result shows that [\textbf{OBLP}] with $n_l = 1$ and $m_l = 1$ is $\mathrm{NP}$-hard.

\begin{lemma}
	\label{lemma 2} The optimistic bilevel linear program 
		\begin{subequations} \label{bilevel min-cost flow}
			\begin{align}
			 z^*_{\mathrm{o}} := 	\max_{x,\mathbf{y}^*}\; 
				& y^*_{\bar e}
			 \\
				\text{\upshape s.t. }
				& x\in[0,\bar x]
			 \\
				& \mathbf{y}^*
				\in
				\argmin_{\mathbf{y}}
				\Big\{
				\sum_{e\in E} c_e y_e :
				\text{\upshape(\ref{cons: min cost flow 1})--(\ref{cons: min cost flow 2})}
				\Big\},
				\label{cons: min-cost flow follower}
			\end{align}
		\end{subequations}
		where $\bar{e} \in E$ denotes a prescribed arc,
		is $\mathrm{NP}$-hard.
		\begin{proof}
	 By Corollary~1.7 in~\cite{Disser2018}, it is $\mathrm{NP}$-hard to decide whether there exists a total flow value $x \in [0,\bar x]$ and an optimal solution of the corresponding parametric min-cost flow problem~(\ref{min cost flow problem}) that sends positive flow through a prescribed arc $\bar e \in E$.
	 This problem is equivalent to deciding whether  $z^*_{\mathrm{o}}$  is strictly positive. Hence, the threshold decision version of~(\ref{bilevel min-cost flow}) is $\mathrm{NP}$-hard.
		\end{proof}
\end{lemma}

Importantly, the follower's problem in~(\ref{cons: min-cost flow follower}) exhibits a very special structure, as the leader variable~$x$ affects only two equality constraints. In the following, we demonstrate that (\ref{bilevel min-cost flow}) reduces to~[\textbf{PBQP}] with a fixed number of follower constraints. 

\begin{theorem} \label{theorem 5}
Under Assumption \textbf{A1}, the pessimistic problem \upshape [\textbf{PBQP}] \itshape with a fixed number of follower constraints $m_f$ is $\mathrm{NP}$-hard, even when $\mathbf{Q}_f \succeq \mathbf{0}$, $\mathbf{P}_l = \mathbf{0}$, $\mathbf{Q}_l = \mathbf{0}$,  and coupling constraints are absent. 
\begin{proof}
First, we note that the follower's feasible region in (\ref{min cost flow problem}) can be expressed as
\begin{subequations} \label{min-cost flow feasible set}
\begin{align}	
	& \mathbf{r} = \mathbf{u} - \mathbf{y} \label{cons: min-cost flow feasible set 1}\\
	& \mathbf{B} \mathbf{y} = 0 \label{cons: min-cost flow feasible set 2}\\
	& \mathbf{b}^{(s)} \mathbf{y} = x \label{cons: min-cost flow feasible set 3} \\
	& \mathbf{b}^{(t)} \mathbf{y} = -x \label{cons: min-cost flow feasible set 4}\\
	& \mathbf{y} \geq \mathbf{0}, \; \mathbf{r} \geq \mathbf{0}, \label{cons: min-cost flow feasible set 5} 
\end{align}
\end{subequations}
where $\mathbf{r}\in\mathbb{R}^{|E|}_+$ is the vector of slack variables associated with the capacity constraints, $\mathbf{b}^{(s)} \in \mathbb{R}^{|E|}$ and $\mathbf{b}^{(t)}\in\mathbb{R}^{|E|}$ are the rows of the node-arc incidence matrix corresponding to $s$ and $t$, respectively, and $\mathbf{B} \in \mathbb{R}^{(|V| - 2) \times |E|}$ is the submatrix corresponding to the intermediate nodes $v\in V\setminus\{s,t\}$. Thus,~(\ref{min-cost flow feasible set}) coincides with the set of optimal solutions of the following convex quadratic~problem:
\begin{subequations} \label{min-cost flow convex quadratic}
	\begin{align}
	\min_{\mathbf{y}, \mathbf{r}} \; & \Vert \mathbf{B} \mathbf{y} \Vert^2 + \Vert \mathbf{r} + \mathbf{y} - \mathbf{u} \Vert^2 \\
	\text{s.t. }	
		&  \text{(\ref{cons: min-cost flow feasible set 3})--(\ref{cons: min-cost flow feasible set 5})}, \label{cons: min-cost flow quadratic 1}  \\ 
		& \sum_{e \in E} (y_e + r_e) \leq \sum_{e \in E} u_e. \label{cons: min-cost flow quadratic 4}
	\end{align}
\end{subequations}
Here, the auxiliary constraint (\ref{cons: min-cost flow quadratic 4}) does not alter the set of optimal solutions in (\ref{min-cost flow convex quadratic}) and ensures that the feasible region (\ref{cons: min-cost flow quadratic 1})--(\ref{cons: min-cost flow quadratic 4}) remains bounded; recall Assumption \textbf{A1}. Furthermore, the optimal objective function value of (\ref{min-cost flow convex quadratic}) is zero whenever the feasible set (\ref{cons: min cost flow 1})--(\ref{cons: min cost flow 2}) is nonempty.

Given an instance of the optimistic problem (\ref{bilevel min-cost flow}), we may assume without loss of generality that~$\mathbf{c} \in\nobreak\mathbb{Z}^{|E|}$. 
For fixed $x \in [0, \bar{x}]$, let $Y(x)$ and $R(x)$ denote  the follower's feasible and reaction sets in (\ref{bilevel min-cost flow}), respectively.   We also define
\[
g(x) := \max \big\{ y_{\bar{e}} : \mathbf{y} \in R(x) \big\}
\]
and the perturbed vector $\tilde{\mathbf{c}}$ such that
$\tilde{c}_{\bar{e}} = c_{\bar{e}} - \tfrac{1}{2}$ and
$\tilde{c}_e = c_e$ for each $e \in E \setminus \{\bar{e}\}$.
Next, we demonstrate~that 
\begin{equation} \label{eq: perturbed min cost flow}
	\min_{\mathbf{y} \in Y(x)} \tilde{\mathbf{c}}^{\top} \mathbf{y} = \min_{\mathbf{y} \in Y(x)} \mathbf{c}^{\top} \mathbf{y} - \frac{1}{2} g(x).
\end{equation}

Let
$\hat{\mathbf{y}}
\in
\argmax_{\mathbf{y} \in R(x)} y_{\bar{e}}$. 
Since $\hat{\mathbf{y}}$ is a minimum-cost flow, i.e., $\hat{\mathbf{y}} \in R(x)$, every simple directed cycle in its residual network has nonnegative cost. Moreover, since all arc costs are integral, every such cycle has either zero cost or cost at least one. Replacing $c_{\bar{e}}$ by $c_{\bar{e}} - \tfrac{1}{2}$ changes the cost of any simple residual cycle by at most~$\tfrac{1}{2}$. Hence, every cycle of positive original cost remains positive after the~perturbation.

Consider now a zero-cost residual cycle. Such a cycle can become negative under the perturbed costs only if it contains the forward residual arc corresponding to $\bar{e}$ but not the backward one. Augmenting along such a cycle yields another optimal flow in $R(x)$ with a strictly larger value of~$y_{\bar{e}}$, contradicting the choice of $\hat{\mathbf{y}}$. Thus, the residual network of $\hat{\mathbf{y}}$ contains no negative-cost cycle under the perturbed costs. Consequently, $\hat{\mathbf{y}}$ is optimal for the perturbed min-cost flow problem, and (\ref{eq: perturbed min cost flow}) holds by the definition of $\tilde{\mathbf{c}}$. 

Consider now the following instance of [\textbf{PBQP}] without coupling constraints:
\begin{subequations} \label{bilevel min-cost flow 2}
	\begin{align}
		z^*_{\mathrm{p}} := \min_{x,\mathbf{y},\mathbf{r}}\;
		& \max_{(\mathbf{y}^*,\mathbf{r}^*)\in\tilde{R}(x)}
		\left\{
		\tilde{\mathbf{c}}^\top\mathbf{y}
		-
		\mathbf{c}^\top\mathbf{y}^*
		\right\}
		\\
		\text{s.t. }
		& \text{(\ref{cons: min-cost flow feasible set 1})--(\ref{cons: min-cost flow feasible set 5})}, \\
		& x\in[0,\bar x],
	\end{align}
\end{subequations}
where
\[
\tilde{R}(x) = \argmin_{\,\mathbf{y}, \mathbf{r}}
\Big\{
\Vert \mathbf{B} \mathbf{y} \Vert^2
+
\Vert \mathbf{r} + \mathbf{y} - \mathbf{u} \Vert^2:
\text{\upshape(\ref{cons: min-cost flow quadratic 1})--(\ref{cons: min-cost flow quadratic 4})}
\Big\}.
\]
By construction, the projection of $\tilde{R}(x)$ onto the $\mathbf{y}$-variables coincides with $Y(x)$. Hence, for every fixed $x \in [0, \bar{x}]$, the optimal value of
(\ref{bilevel min-cost flow 2}) satisfies
\[
\min_{\mathbf{y} \in Y(x)}
\tilde{\mathbf{c}}^\top\mathbf{y}
- \min_{\mathbf{y} \in Y(x)}
\mathbf{c}^\top\mathbf{y} = -\tfrac{1}{2}g(x),
\]
where the equality follows from
(\ref{eq: perturbed min cost flow}). Consequently, the optimal objective function value of (\ref{bilevel min-cost flow 2}) is
\[
z^*_{\mathrm{p}} = -\tfrac{1}{2}\max_{x\in[0,\bar x]}g(x) = -\tfrac{1}{2} z^*_{\mathrm{o}}.
\]
Hence, by Lemma~\ref{lemma 2}, deciding whether $z^*_{\mathrm{p}} < 0$ is $\mathrm{NP}$-hard.

Finally, note that $(x, \mathbf{y}, \mathbf{r})$ in (\ref{bilevel min-cost flow 2}) represent leader variables, and the follower's problem (\ref{min-cost flow convex quadratic}) has nonnegativity constraints (\ref{cons: min-cost flow feasible set 5}), two equality constraints (\ref{cons: min-cost flow feasible set 3})--(\ref{cons: min-cost flow feasible set 4}) and inequality constraint~(\ref{cons: min-cost flow quadratic 4}). Since each equality constraint can be replaced by two inequalities, we thus obtain an instance of~[\textbf{PBQP}] with~$m_f = 5$.

By taking $\bar x$ to be the maximum $s$-$t$ flow value in the constructed network, we conclude that the follower's feasible set (\ref{cons: min-cost flow quadratic 1})--(\ref{cons: min-cost flow quadratic 4}) is nonempty and bounded for every $x\in[0,\bar x]$. The leader variables are also bounded because $x\in[0,\bar x]$, $\mathbf{r} \geq \mathbf{0}$, $\mathbf{y} \geq \mathbf{0}$, and $\mathbf{r} + \mathbf{y} = \mathbf{u}$. Hence, the constructed instance satisfies Assumption~\textbf{A1}, and the result follows. 
\end{proof} 
\end{theorem}

Next, we show that the polynomial-time solvability result of Theorem~\ref{theorem 4} no longer holds when the follower's objective function $z_f(\mathbf{y})$ is allowed to be nonconvex, i.e., when $\mathbf{Q}_f$ is indefinite. For this reduction, we use a bilinear optimization problem
\begin{subequations} \label{product problem}
	\begin{align}
		\min_{\mathbf{x}} \; & x_1 x_2 \\
		\mbox{s.t. } & \mathbf{A} \mathbf{x} \leq \mathbf{b} \label{cons: product problem 1} \\
		& \mathbf{x} \in [0,1]^n, \label{cons: product problem 2}
	\end{align}
\end{subequations}
which is known to be $\mathrm{NP}$-hard~\cite{Matsui1996}. Here, we assume that the feasible set~(\ref{cons: product problem 1})--(\ref{cons: product problem 2}) is nonempty. Moreover, since the hardness result in~\cite{Matsui1996} holds for bounded polyhedral feasible regions with~$\mathbf{x} \geq \mathbf{0}$, we may assume without loss of generality that $\mathbf{x} \in [0,1]^n$. The following result holds.

\begin{theorem} \label{theorem 6}
Under Assumption \textbf{A1}, the optimistic problem \upshape [\textbf{OBQP}] \itshape with fixed numbers of follower variables and constraints, $n_f$ and $m_f$, is $\mathrm{NP}$-hard, even when $\mathbf{P}_l = \mathbf{0}$, $\mathbf{Q}_l = \mathbf{0}$, and $\mathbf{Q}_f$ is indefinite.
\begin{proof}
Consider the following epigraph reformulation of (\ref{product problem}): 
\begin{subequations} \label{product problem 2}
	\begin{align}
		\min_{\mathbf{x}, t} \; & t \\
		\mbox{s.t. } & \text{(\ref{cons: product problem 1})--(\ref{cons: product problem 2})}, \\
		& t \in [0, 1] \\
		& x_1 x_2 \leq t, \label{cons: product}
	\end{align}
\end{subequations}
where we assume that $t \in [0,1]$ without loss of generality. 

The idea is then to enforce the nonconvex constraint (\ref{cons: product}) using the follower's problem (\ref{cons: obqp follower}) and optimistic coupling constraints (\ref{cons: obqp leader}). In this regard, we introduce the following nonconvex quadratic follower's problem:
\begin{subequations} \label{follower's problem nonconvex}
	\begin{align}
		\min_{\mathbf{y}, s} \; & -y_1 y_2 + sy_3 + s(1 - s) \\
		\mbox{s.t. } & y_3 = t \label{cons: follower's problem nonconvex 1} \\
		& 0 \leq y_i \leq x_i \quad \forall i \in \{1, 2\} \label{cons: follower's problem nonconvex 2} \\
		& y_i \leq s \quad \forall i \in \{1, 2\} \label{cons: follower's problem nonconvex 3} \\
		& s \in [0, 1], \quad y_3 \geq 0. \label{cons: follower's problem nonconvex 4} 
	\end{align}
\end{subequations}
and a coupling constraint $s^* \leq 0$. 

First, observe that if a selected optimal solution of (\ref{follower's problem nonconvex}) satisfies the coupling constraint $s^*\leq 0$, then, since $s\in[0,1]$, we must have $s^*=0$. Consequently, constraints~(\ref{cons: follower's problem nonconvex 1})--(\ref{cons: follower's problem nonconvex 4}) imply $\mathbf{y}^*=(0,0,t)^\top$, and the corresponding follower's objective function value is~$0$. 

Next, we consider two possible cases:
\begin{itemize}
	\item Assume that $x_1 x_2 > t$. By setting $\tilde{y}^* = (x_1, x_2, t)^\top$ and $\tilde{s}^* = 1$, we observe that
	\[-\tilde{y}^*_1 \tilde{y}^*_2 + \tilde{s}^* \tilde{y}^*_3 + \tilde{s}^*(1 - \tilde{s}^*) = -x_1 x_2 + t < 0,\]
 and thus the optimal objective function value of (\ref{follower's problem nonconvex}) is strictly negative. Consequently, optimal solutions of (\ref{follower's problem nonconvex}) do not satisfy the coupling constraint $s^* \leq 0$.
 \item Assume that $x_1 x_2 \leq t$. We note that the follower's problem (\ref{follower's problem nonconvex}) reduces to:
 \begin{equation} \label{follower's problem nonconvex 2}
 		\min_{s} \; \Big\{ -\min\{x_1, s\} \min\{x_2, s\} + st + s(1- s): s \in [0, 1] \Big\}.
 \end{equation}
 Furthermore, the objective function in (\ref{follower's problem nonconvex 2}) is piecewise concave in $s$. Hence, its minimum is obtained at one of the breakpoints, i.e., $s^* \in \{0, x_1, x_2, 1\}$. Without loss of generality, let $0 \leq x_1 \leq x_2 \leq 1$. Then, the optimal objective function value in (\ref{follower's problem nonconvex 2}) is given by
 \begin{equation} \label{follower's problem nonconvex opt}
 	 \min\Big\{0, x_1(t + 1 - 2x_1), x_2(t - x_1 + 1 - x_2), -x_1 x_2 + t \Big\} = 0.
 \end{equation} 
 Indeed, by using $x_1 x_2 \leq t$ and $0 \leq x_1 \leq x_2 \leq 1$, we observe that
 \begin{align*}
 	& x_1(t + 1 - 2x_1) \geq x_1(x^2_1 + 1 - 2x_1) = x_1(x_1 - 1)^2 \geq 0, \\
 	& x_2(t - x_1 + 1 - x_2) \geq x_2(x_1 x_2 - x_1 + 1 - x_2) = x_2(1 - x_2)(1 - x_1) \geq 0,
 \end{align*}
 and thus the equality in (\ref{follower's problem nonconvex opt}) holds. Consequently, there exists an optimal solution of (\ref{follower's problem nonconvex}), namely $\mathbf{y}^*=(0,0,t)^\top$ and $s^* = 0$, that satisfies $s^* \leq 0$. 
 
\end{itemize}

We conclude that constraint~(\ref{cons: product}) is satisfied if and only if there exists an optimal solution of~(\ref{follower's problem nonconvex}) satisfying $s^* \leq 0$. As a result, (\ref{product problem 2}) reduces to the following instance of~[\textbf{OBQP}]: 
\begin{subequations} \label{product problem 3}
	\begin{align}
		\min_{\mathbf{x}, t, \mathbf{y}^*, s^*} \; & t \\
		\mbox{s.t. } & \text{(\ref{cons: product problem 1})--(\ref{cons: product problem 2})}, \\
		& t \in [0, 1] \\
		& s^* \leq 0 \\
		& (\mathbf{y}^*, s^*) \in \argmin_{\,\mathbf{y}, s} \Big\{-y_1 y_2 + sy_3 + s(1 - s): \text{(\ref{cons: follower's problem nonconvex 1})--(\ref{cons: follower's problem nonconvex 4})}\Big\}. \label{cons: product problem 3 follower}
	\end{align}
\end{subequations}
By design, Assumption \textbf{A1} is satisfied and the follower's problem in (\ref{cons: product problem 3 follower}) has $n_f = 4$ variables and~$m_f = 7$ constraints, with the equality constraint (\ref{cons: follower's problem nonconvex 1}) being replaced by two inequalities. This observation concludes the proof. 
\end{proof}
\end{theorem}

\section{Conclusion} \label{sec: conclusion}
 \looseness-1 Overall, our results identify structural mechanisms of computational hardness in bilevel linear programs (BLPs) and bilevel quadratic programs (BQPs) under restrictions on the numbers of variables or constraints. First, we demonstrate that, under comparable assumptions, \textit{pessimistic} BLPs can be computationally harder than their \textit{optimistic} counterparts, with this complexity transition driven by the presence of \textit{coupling constraints}. We further show that the \textit{form} of the follower's objective function constitutes an alternative source of hardness. Thus, replacing a linear follower objective with a convex quadratic one is sufficient to induce hardness in the pessimistic setting even in the absence of coupling constraints. Finally, allowing a nonconvex quadratic follower objective leads to a further complexity transition in the optimistic setting, even when both follower dimensions are fixed.    

Several questions remain open for future research. First, the complexity of optimistic bilevel programs with a convex quadratic follower objective and a fixed number of follower constraints remains unresolved. Second, the $\mathrm{NP}$-hardness results for bilevel quadratic programs provide lower complexity bounds only. Establishing matching upper bounds and determining the exact complexity of these problem classes constitute important directions for future research.

\onehalfspacing
 \bibliographystyle{apa}
 \bibliography{bibliography}
\end{document}